\documentclass[english]{article}
\usepackage[T1]{fontenc}
\usepackage[latin9]{luainputenc}
\usepackage{geometry}
\usepackage{amstext}
\usepackage{amssymb}
\usepackage{feyn}
\usepackage{tikz}
\usepackage{graphicx}
\usepackage{array}
\usepackage{booktabs}

\usepackage{amsmath}
\usepackage{amsfonts}

\usepackage[normalem]{ulem}
\usepackage{color}
\usepackage{listings,braket}
\usepackage{caption}
\usepackage{subcaption}
\usepackage{float}
\definecolor{darkgreen}{rgb}{0,0.35,0}
\definecolor{Rood}{rgb}{1, 0, 0}

\makeatletter
\usepackage[english]{babel}
\usepackage{feyn}

\makeatother

\begin{document}

\title{\textbf{On the representations of Bell's operators in Quantum Mechanics }}

{\author{\textbf{S.~P.~Sorella}\\\\\
\textit{{\small $^1$UERJ -- Universidade do Estado do Rio de Janeiro,}}\\
\textit{{\small Instituto de F\'{\i}sica -- Departamento de F\'{\i}sica Te\'orica -- Rua S\~ao Francisco Xavier 524,}}\\
\textit{{\small 20550-013, Maracan\~a, Rio de Janeiro, Brasil}}\\
}

\date{}

\maketitle
\begin{abstract}
We point out that, when the dimension of the Hilbert space is greater than two, Bell's operators entering the Bell-CHSH inequality  exhibit unitarily inequivalent representations. Although the Bell-CHSH inequality turns out to be violated, the size of the violation is different for different representations, the maximum violation being given by Tsirelson's bound. The feature relies on a pairing mechanism between the modes  of the Hilbert space of the system. 
\end{abstract}

\section{Introduction}\label{intro}
The Bell-CHSH inequality \cite{Bell:1964kc,Clauser:1969ny} is one of the most fundamental topics of  Quantum Mechanics, signaling  the existence of very strong  correlations, due to the phenomenon of entanglement. \\\\In its usual form, it reads 
\begin{equation}
 \langle \psi | {\cal C}_{CHSH} | \psi \rangle  =  \langle 
 \psi  (A_1+A_2)B_1 + (A_1-A_2)B_2 | \psi \rangle     \;,  \label{BCHSH}
\end{equation}
where $| \psi \rangle$ is a given entangled state and $(A_i, B_k)$, $i,k=1,2$, are a set of four Hermitian operators fulfilling the  requirements \cite{tsi1}: 
\begin{equation} 
A^2_i=1 \;, \qquad B^2_k =1 \;, \qquad [A_i, B_k] = 0 \;. \label{req}
\end{equation}
One speaks of a violation of the Bell-CHSH inequality whenever 
\begin{equation} 
2 < |  \langle \psi | {\cal C}_{CHSH} | \psi \rangle | \le 2\sqrt{2} \;. \label{vb}
\end{equation}
where the maximum violation, $2\sqrt{2}$, is known as  Tsirelson's  bound \cite{tsi1}. As the notation itself let it understand, expression \eqref{BCHSH} refers to a bipartite system $AB$ whose Hilbert space is ${\cal H} = {\cal H}_a \otimes {\cal H}_b$, with  ${\cal H}_a$  and ${\cal H}_b$ having the same dimension:  $d_a = d_b=d$. The operators $A_i$ act only on ${\cal H}_a$, while $B_k$  only on ${\cal H}_b$. \\\\Up to unitary transformations, the construction of the four operators $(A_i, B_k)$ is unique when  $d=2$. This is the original example of spin $1/2$ particles considered by Bell \cite{Bell:1964kc}, where $(A_i, B_k)$ are expressed in terms of Pauli matrices. \\\\Though, this is no more true when the Hilbert spaces ${\cal H}_a$ and  ${\cal H}_b$ have dimension greater than two. It is the aim of the present work to show that in such cases there exist unitarily inequivalent representations which lead to different values of the violation of the Bell-CHSH inequality. The number of such inequivalent representations becomes larger and larger as $d$ increases. \\\\This observation is based on the well known work by \cite{popr}, see Theorem 3. More precisely, the aforementioned inequivalent representations turn out to be linked to the presence of the two by two matrix \begin{equation}
{\cal M}_{i} = {\cal M}_{i}^{\dagger } = 
\begin{pmatrix}
0 & e^{i \alpha_i}  \\
e^{-i \alpha_i}  & 0  \\
\end{pmatrix} 
\;, \label{Bmx}
\end{equation} 
which will enable us to introduce a pairing mechanism between the various modes of the Hilbert space. As we shall see, the inequivalent representations can be labelled according to the number of times in which the matrix ${\cal M}_{i}$ appears. In particular, the representation fully given by the direct sum of ${\cal M}_{i}$ yields the maximum allowed violation, namely: Tsirelson's bound $2\sqrt{2}$.

\section{The example of a four dimensional Hilbert space} \label{four}

We start by considering the case in which ${\cal H}_a$ and  $ {\cal H}_b$ have dimension four. Let $(| 0 \rangle_a, | 1 \rangle_a, | 2 \rangle_a, | 3 \rangle_a)$ and $(| 0 \rangle_b, | 1 \rangle_b, | 2 \rangle_b, | 3 \rangle_b)$ denote  orthonormal basis of  ${\cal H}_a$, resp. $ {\cal H}_b$. Both operators $(A_i, B_k)$ can be represented by $4\times 4$ matrices. As entangled state, we shall consider the maximally entangled state
\begin{equation} 
| \psi \rangle = \frac{ | 0 \rangle_a | 0 \rangle_b + | 1 \rangle_a | 1 \rangle_b + | 2 \rangle_a | 2 \rangle_b + | 3 \rangle_a | 3 \rangle_b}{\sqrt{4}} \;. \label{state4}
\end{equation}
To construct the operators $A_i$, resp. $B_k$,  one has to keep in mind that they have to be defined on the whole Hilbert space ${\cal H}_a$, resp. $ {\cal H}_b$
\subsection{First representation} 
A first possibility for $(A_i, B_k)$ is given by 
\begin{eqnarray} 
A_i | 0 \rangle_a  =  | 0  \rangle_a \;, \qquad A_i | 1 \rangle_a = e^{i \alpha_i} | 2 \rangle_a \;, \qquad  A_i | 2 \rangle_a  =  e^{-i \alpha_i} | 1  \rangle_a \;, \qquad A_i | 3 \rangle_a  =  | 3  \rangle_a \;, \nonumber \\[2mm]
B_k | 0 \rangle_b  =  | 0  \rangle_b \;, \qquad B_k | 1 \rangle_b = e^{i \beta_k} | 2 \rangle_b \;, \qquad  B_k | 2 \rangle_b  =  e^{-i \beta_k} | 1  \rangle_b \;, \qquad B_k | 3 \rangle_b  =  | 3  \rangle_b \;,  \label{firstAB}
\end{eqnarray} 
where $(\alpha_i, \beta_k)$ stand for arbitrary real parameters. The operators $(A_i, B_k)$ defined in that way are Hermitian and fulfill the requirements \eqref{req}. \\\\A quick computation shows that 
\begin{equation} 
  A_i B_k |\psi \rangle = \frac{  | 0 \rangle_a | 0 \rangle_b + e^{-i(\alpha_i+\beta_k)} | 1 \rangle_a | 1 \rangle_b + 
  e^{i(\alpha_i +\beta_k)} | 2 \rangle_a | 2 \rangle_b + | 3 \rangle_a | 3 \rangle_b}{\sqrt{4}} \;, \label{state4AB}
\end{equation}
so that 
\begin{equation}
\langle \psi |  A_i B_k |\psi \rangle = \frac{ \cos(\alpha_i + \beta_k) + 1}{2} \;.  \label{AB4}
\end{equation} 
Therefore, for the Bell-CHSH combination, we get 
\begin{equation} 
 \langle \psi | {\cal C}_{CHSH} | \psi \rangle =  \frac{ \cos(\alpha_1 + \beta_1) + \cos(\alpha_2 + \beta_1)+ \cos(\alpha_1 + \beta_2) - \cos(\alpha_2 + \beta_2)+ 2}{2}. \label{CHSH14}
\end{equation}
To obtain maximum violation, one sets \cite{tsi1}
\begin{equation} 
\cos(\alpha_1 + \beta_1) + \cos(\alpha_2 + \beta_1)+ \cos(\alpha_1 + \beta_2) - \cos(\alpha_2 + \beta_2) = 2 \sqrt{2} \;, \label{max14}
\end{equation} 
namely
\begin{equation} 
\alpha_1 = 0 \;, \qquad \alpha_2 = \frac{\pi}{2} \;, \qquad \beta_1= -\frac{\pi}{4} \;, \qquad \beta_2=  \frac{\pi}{4} \;, \label{choice14} 
\end{equation} 
resulting in 
\begin{equation} 
 \langle \psi | {\cal C}_{CHSH} | \psi \rangle = \sqrt{2} + 1 \approx 2.4 \label{CHSH14res}
\end{equation}
We underline that the above value is the maximum violation allowed by the definition \eqref{firstAB}. Looking at the matrix representation for $(A_i, B_k)$, one gets 
\begin{equation}
A_{i}^{(1)} = A_{i}^{(1)\dagger} =
\begin{pmatrix}
1 & 0  & 0 & 0 \\
0  & 0 & e^{i\alpha_i} & 0 \\
0  & e^{-i\alpha_i}   & 0 & 0  \\
0 & 0 & 0 & 1
\end{pmatrix}  
 \;, \label{opAB14}
\end{equation}
and similarly for $B_k^{(1)}$, with $\alpha_i$ replaced by $\beta_k$.  
\subsection{Second representation}
A second representation might be obtained by following the setup outlined in \cite{Summ}  in the study of the violation of the Bell-CHSH inequality in relativistic Quantum Field Theory. One defines 
\begin{eqnarray} 
A_i | 0 \rangle_a  =  e^{i \alpha_i} | 1  \rangle_a \;, \qquad A_i | 1 \rangle_a = e^{-i \alpha_i} | 0 \rangle_a \;, \qquad  A_i | 2 \rangle_a  =  e^{i \alpha_i} | 3  \rangle_a \;, \qquad A_i | 3 \rangle_a  = e^{-i\alpha_i}  | 2  \rangle_a \;, \nonumber \\[2mm]
B_k | 0 \rangle_b  = e^{i \beta_k} | 1  \rangle_b \;, \qquad B_k | 1 \rangle_b = e^{-i \beta_k} | 0 \rangle_b \;, \qquad  B_k | 2 \rangle_b  =  e^{i \beta_k} | 3  \rangle_b \;, \qquad B_k | 3 \rangle_b  = e^{-i\beta_k} | 2  \rangle_b \;,  \label{secondAB}
\end{eqnarray} 
Again, the operators are Hermitian and fulfill conditions \eqref{req}. For the correlation function  $\langle \psi |  A_i B_k |\psi \rangle $, one gets
\begin{equation}
\langle \psi |  A_i B_k |\psi \rangle = \cos(\alpha_i + \beta_k)  \;.  \label{secAB4}
\end{equation} 
so that  
\begin{equation} 
 \langle \psi | {\cal C}_{CHSH} | \psi \rangle =  \left( \cos(\alpha_1 + \beta_1) + \cos(\alpha_2 + \beta_1)+ \cos(\alpha_1 + \beta_2) - \cos(\alpha_2 + \beta_2) \right). \label{secCHSH14}
\end{equation}
Making use of the choice \eqref{choice14}, one finds maximum violation, {\it i.e.} the saturation of Tsirelson's bound: 
\begin{equation} 
 \langle \psi | {\cal C}_{CHSH} | \psi \rangle = 2 \sqrt{2} \;. \label{CHSH24res}
\end{equation}
In matrix representation, we have now 
 \begin{equation}
A_{i}^{(2)} = A_{i}^{(2)\dagger} =
\begin{pmatrix}
0 & e^{i\alpha_i}  & 0 & 0 \\
e^{-i \alpha_i}  & 0 & 0 & 0 \\
0  & 0   & 0 & e^{i \alpha_i}   \\
0 & 0 & e^{-i \alpha_i} & 0
\end{pmatrix}  
 \;. \label{opAB24}
\end{equation}
However, from Specht's theorem, it follows that $A_{i}^{(1)}$ and $A_{i}^{(2)}$ cannot be unitarily equivalent, since
\begin{equation} 
{\rm Tr} A_{i}^{(1)} =2 \neq {\rm Tr } A_{i}^{(2)} =0 \;. \label{inquiv4}
\end{equation} 

One observes that the maximum violation, {\it i.e.} the saturation of Tsirelson's bound, occurs for the representation for which ${\rm Tr} A_{i}=0$, namely eq.\eqref{opAB24}. This feature persists for all examples which will be analyzed. As we shall see,  the condition ${\rm Tr} A_{i}=0$ can be seen as a consequence of the fact that this representation is  made up by the direct sum of the 2x2 Pauli matrices employed for discussing the violation of the Bell-CHSH inequality for the two-dimensional Hilbert space of spin $1/2$.  

\subsection{The case of a non-maximally entangled state} 
So far, we have considered only the case of maximally entangled states. It is instructive to check out what happens if the starting state is not maximally entangled as. for instance:
\begin{equation} 
| \psi \rangle = \frac{ | 0 \rangle_a | 0 \rangle_b + | 1 \rangle_a | 1 \rangle_b + | 2 \rangle_a | 2 \rangle_b + \sqrt{r} | 3 \rangle_a | 3 \rangle_b}{\sqrt{3+r}} \;. \label{state4r}
\end{equation}
where $r$ is a real parameter $0 < r < 1$. Let us discuss the violation for each representation: 
\begin{itemize} 
 
\item first representation, eq.\eqref{opAB14}. \\\\A simple calculation shows that 
\begin{equation} 
 \langle \psi | {\cal C}_{CHSH} | \psi \rangle =  \frac{2(1+r)}{3+r} + \frac{4 \sqrt{2}}{3+r}  \;. \label{firstnm} 
\end{equation}

\item second representation, eq.\eqref{secondAB}. \\\\It turns out that  
\begin{equation} 
 \langle \psi | {\cal C}_{CHSH} | \psi \rangle =  4\sqrt{2} \left(\frac{\sqrt{r}}{3+r}\right) + \frac{4 \sqrt{2}}{3+r}\;. \label{CHSHnmr}
\end{equation}
Since the parameter $r$ is always $r<1$, both expressions \eqref{firstnm}, \eqref{CHSHnmr} cannot saturate Tsirelson's bound: $2\sqrt{2}$. In the first case, eq.\eqref{firstnm}, the maximum violation is attained for $r \approx 0$, yielding $\approx 2.55$. In the second case, the violation is bigger, increasing as $r$ approximates the value $1$.  Tsirelson bound is attained when $r=1$, {\it i.e.} when the state  \eqref{state4r} becomes maximally entangled.  \\\\Other kinds of non-maximally entangled states might be discussed. Though, we shall proceed by considering examples of maximally entangled states. 

\end{itemize}

\section{The six dimensional case}\label{six}
For a better understanding, let us discuss briefly the six dimensional case, considering the maximally entangled state
\begin{equation} 
| \psi \rangle = \frac{ | 0 \rangle_a | 0 \rangle_b + | 1 \rangle_a | 1 \rangle_b + | 2 \rangle_a | 2 \rangle_b + | 3 \rangle_a | 3 \rangle_b+  | 4 \rangle_a | 4 \rangle_b+  | 5 \rangle_a | 5 \rangle_b}{\sqrt{6}} \;. \label{state6}
\end{equation}
As the dimension of the Hilbert state increases, more inequivalent matrix representations can be constructed, namely 
\begin{itemize} 
\item first option 
 \begin{equation}
A_{i}^{(1)} = A_{i}^{(1)\dagger} =
\begin{pmatrix}
1 & 0 & 0 &0 & 0 & 0 \\
0 & 0 & e^{i\alpha_i}  & 0 & 0 & 0 \\
0& e^{-i \alpha_i}  & 0 & 0 & 0 & 0\\
0  & 0   & 0 & 1 & 0 & 0   \\
0 & 0 & 0 & 0 & 1 & 0 \\
0 & 0 & 0 & 0 & 0 & 1 \\
\end{pmatrix}  
\end{equation}
\begin{equation}
A_{i}^{(1)}A_{i}^{(1)} = 1, \qquad {\rm Tr}A_{i}^{(1)} = 4\;, \qquad  \langle \psi | {\cal C}_{CHSH} | \psi \rangle^{(1)} = \frac{2 (2 \sqrt{2}) + 2 \cdot 4}{6} \approx 2.27
 \;, \label{opAB16}
 \end{equation}
\item second option 
 \begin{equation}
A_{i}^{(2)} = A_{i}^{(2)\dagger} =
\begin{pmatrix}
1 & 0 & 0 &0 & 0 & 0 \\
0 & 0 & e^{i\alpha_i}  & 0 & 0 & 0 \\
0& e^{-i \alpha_i}  & 0 & 0 & 0 & 0\\
0  & 0  & 0 & 0 & e^{i \alpha_i} & 0   \\
0 & 0 & 0 & e^{-i\alpha_i}  & 0 & 0 \\
0 & 0 & 0 & 0 & 0 & 1 \\
\end{pmatrix}  
\end{equation}
\begin{equation}
A_{i}^{(2)}A_{i}^{(2)} = 1, \qquad {\rm Tr}A_{i}^{(2)} = 2\;, \qquad  \langle \psi | {\cal C}_{CHSH} | \psi \rangle^{(2)} = \frac{4 (2 \sqrt{2} )+  4}{6} \approx 2.53
 \;, \label{opAB26}
 \end{equation}
 \item third option 
 \begin{equation}\label{sixmatr}
A_{i}^{(3)} = A_{i}^{(3)\dagger} =
\begin{pmatrix}
0 & e^{i \alpha_i} & 0 &0 & 0 & 0 \\
e^{-i \alpha_i}  & 0 & 0  & 0 & 0 & 0 \\
0& 0  & 0 & e^{i \alpha_i} & 0 & 0\\
0  & 0  & e^{-i \alpha_i}  & 0 & 0 & 0   \\
0 & 0 & 0 & 0  & 0 & e^{i \alpha_i} \\
0 & 0 & 0 & 0 & e^{-i \alpha_i} & 0 \\
\end{pmatrix}  
\end{equation}
\begin{equation}
A_{i}^{(3)}A_{i}^{(3)} = 1, \qquad {\rm Tr}A_{i}^{(3)} = 0\;, \qquad  \langle \psi | {\cal C}_{CHSH} | \psi \rangle^{(3)} = \frac{(2+2+2) (2 \sqrt{2}) }{6} = 2 \sqrt{2} 
 \;, \label{opAB36}
 \end{equation}
 \end{itemize} 
 All three representations are unitarily inequivalent, yielding different violations for the Bell-CHSH inequality. \\\\ We notice that, once again, the maximum violation occurs when ${\rm Tr}A_{i} = 0$. 
 
\section{A general pattern: even and odd finite dimensional Hilbert spaces}\label{general}
A general pattern emerges from the previous analysis. Let us consider the case in which both 
${\cal H}_a$ and  $ {\cal H}_b$ have generic finite dimension $N$.  As entangled state, we shall take  the maximally entangled state 
\begin{equation} 
| \psi \rangle = \frac{1}{\sqrt{N}} \sum_{n=0}^{N-1} |n\rangle_a |n \rangle_b. \;. \label{MNst}
\end{equation} 
The pattern is encoded in the action of the two by two matrix 
\begin{equation}
{\cal M}_{i} = {\cal M}_{i}^{\dagger } = 
\begin{pmatrix}
0 & e^{i \alpha_i}  \\
e^{-i \alpha_i}  & 0  \\
\end{pmatrix} 
\;, \label{MB}
\end{equation} 
which is nothing but Bell's original operator \cite{Bell:1964kc}, namely 
\begin{equation} 
{\cal M}_{i} = {\vec n}_i \cdot {\vec \sigma}   \label{nsigma}
\end{equation} 
where $\vec \sigma$ are the Pauli matrices and ${\vec n}_i$ is the unit vector ${\vec n}_i= (\cos(\alpha_i), -\sin(\alpha_i), 0)$. \\\\Two orthogonal modes $|\xi_a\rangle$ and $| \chi_a\rangle$, $\langle \xi_a | \chi_a \rangle=0$, entering the state \eqref{MNst} are said to form a pair if 
\begin{equation} 
A_i 
\begin{pmatrix} 
\xi_a \\
\chi_a 
\end{pmatrix}
= {\cal M}_{i}
\begin{pmatrix} 
\xi_a \\
\chi_a 
\end{pmatrix} 
\;, \label{pair}
\end {equation} 
{\it i.e.} 
\begin{equation}
A_i | \xi_a \rangle = e^{i \alpha_i} | \chi_a \rangle \;, \qquad  A_i | \chi_a \rangle = e^{-i \alpha_i} | \xi_a \rangle \;. \label{p2pair}
\end{equation}
The matrix ${\cal M}_i$ turns out to be the building block of all matrix representations listed above. Its action is that of forming pairs out of the $N$ modes contributing to the state \eqref{MNst}. Each pair gives a contribution $ 2(2\sqrt{2})$ to the Bell-CHSH inequality, as it is apparent from eqs.\eqref{opAB16},  \eqref{opAB26}, \eqref{opAB36}, depending on how many times the matrix ${\cal M}_{i}$ appears in a given representation. It is thus natural to split the Hilbert spaces in two categories: even and odd dimensional Hilbert spaces. \\\\In the case in which $N$ is even, we can have inequivalent representations in which the appearance of the matrix  ${\cal M}_{i}$ ranges from $1$ to $N/2$. Since the maximum number of pairs is $N/2$, we immediately get that the violation of the Bell-CHSH lies in the interval 
\begin{equation} 
\frac{ 2(2\sqrt{2}) + 2\cdot (N-2)}{N}   \le \langle \psi | {\cal C}_{CHSH} | \psi \rangle \le \frac{ 2\cdot \frac{N}{2} (2\sqrt{2}) }{N}= 2\sqrt{2} \;, \qquad N \ge 4 \;. \label{reven}
\end{equation} 
On the other hand, in the odd case, it is impossible to convert all modes into pairs, so that the size of the Bell-CHSH is always strictly lower than Tsirelson's bound, namely 
\begin{equation} 
\frac{ 2(2\sqrt{2}) + 2\cdot (2N-1)}{2N+1}   \le \langle \psi | {\cal C}_{CHSH} | \psi \rangle \le  = 2\sqrt{2} - \frac{2}{2N+1}\;, \qquad N=1.2,.... \label{revodd}
\end{equation}
We see thus that a representation in which  ${\cal M}_{i}$ appears only once, {\it i.e.} only one pair of modes is contemplated, already leads to a violation of the Bell-CHSH inequality which, however, attains its minimum value. In the even case, Tsirelson's bound is recovered when all modes have been converted into pairs, {\it i.e.} the representation is built out  through the direct sum of $N/2$ matrices ${\cal M}_{i}$. In the odd case, it is impossible to achieve Tsirelson's bound, as a perfect pairing cannot be done. We recover here the result by Gisin-Peres \cite{Gisin}, see also \cite{Peruzzo:2023nrr,Sorella:2023hku}.

\section{The infinite dimensional case: the squeezed state}\label{squeezed}

Let us move now to the infinite dimensional case, by considering the  normalized squeezed state

\begin{equation} 
|\eta \rangle = \sqrt{(1-\eta^2)} \;e^{\eta a^\dagger b^\dagger} |0 \rangle \;, \qquad \langle \eta | \eta \rangle = 1 \;,  \label{etast}
\end{equation} 
where the real parameter $\eta$ is constrained to belong to the interval 
\begin{equation} 
0 < \eta < 1 \;. \label{intv} 
\end{equation} 
The operators $(a,b)$ obey the following commutation relations: 
\begin{eqnarray} 
 \left[ a, a^{\dagger}\right] & = & 1\;, \qquad [a^{\dagger}, a^{\dagger}] =0 \;, \qquad  [a, a] =0 \;, \nonumber \\
 \left[ b, b^{\dagger}\right]  & = & 1\;, \qquad [b^{\dagger}, b^{\dagger}] =0 \;, \qquad  [b, b] =0 \;, \nonumber \\
 \left[a, b \right] & = &  0\;, \qquad [a, b^{\dagger}] =0 \;,  \label{ccrqm}
\end{eqnarray} 
with
\begin{equation} 
a |0\rangle = b |0\rangle = 0 \;. \label{st}
\end{equation}
As one can easily figure out, we have now an infinite number of inequivalent possibilities, each leading to a violation of the Bell-CHSH inequality. As said before, the minimum violation is obtained by considering only one pair of modes. Let us pick up  the modes $|0 \rangle$ and $|1 \rangle$ in expression \eqref{etast}, namely, we write 
\begin{equation} 
|\eta \rangle = \sqrt{(1-\eta^2)} \left( |0_a\rangle |0_b \rangle  + \eta  |1_a\rangle |1_b \rangle  + \sum_{n=2}^{\infty} \eta^n |n_a\rangle |n_b \rangle \right) \;,  \label{etast1}, 
\end{equation} 
where 
\begin{equation} 
|n_a\rangle |n_b \rangle = \frac{1}{n!} (a^\dagger)^n (b^\dagger)^n |0\rangle \;, \qquad |0\rangle = |0_a\rangle|0_b\rangle \;. \label{vc}
\end{equation} 
For the operators  $(A_i, B_k)$, we write 
\begin{eqnarray} 
A_i | 0_a \rangle = e^{i \alpha_i} | 1_a \rangle \;, \qquad A_i | 1_a \rangle = e^{-i \alpha_i} | 0_a \rangle \;, \qquad A_i = 1\;\;  {\it  on \; the\; remaining \; elements \;of \;the \;basis} \;, \nonumber \\[2mm]
B_k | 0_b \rangle = e^{i \beta_k} | 1_b \rangle \;, \qquad B_k | 1_b \rangle = e^{-i \beta_k} | 0_b \rangle \;, \qquad B_k = 1\;\;  {\it  on \; the\; remaining \; elements \;of \;the \;basis } \;. \label{ABopcoh}
\end{eqnarray}
Therefore, 
\begin{equation} 
\langle \eta | A_i B_k | \eta \rangle = 1 + (1-\eta^2) \left(  2 \eta \cos(\alpha_i + \beta_k)  -1 -\eta^2 \right)  \;. \label{etaAB}
\end{equation} 
Setting 
\begin{equation} 
\alpha_1 = 0 \;, \qquad \alpha_2 = \frac{\pi}{2} \;, \qquad \beta_1= -\frac{\pi}{4} \;, \qquad \beta_2=  \frac{\pi}{4} \;, \label{choicesq} 
\end{equation} 
 for the Bell-CHSH inequality one gets 
 \begin{equation}
\langle \eta | {\cal C}_{CHSH} | \eta  \rangle = 2 + 2 (1-\eta^2) \left( 2 \sqrt{2}\; \eta -1 -\eta^2 \right)\;. \label{cohBCHSHsq} 
\end{equation}
 There is violation whenever 
 \begin{equation} 
 \sqrt{2} - 1 < \eta < 1 \;. \label{vsq}
 \end{equation} 
 The maximum value of the violation occurs for $\eta \approx 0.7$, yielding 
 \begin{equation}
\langle \eta | {\cal C}_{CHSH} | \eta  \rangle \approx 2.5\;. \label{cohBCHSHsqv} 
\end{equation} 
On the other hand, the maximum violation is obtained when the pairing mechanism involves all modes, namely: 
\begin{eqnarray} 
A_i | (2n)_a \rangle &=& e^{i \alpha_i} | (2n+1)_a \rangle \;, \qquad A_i | (2n+1)_a \rangle = e^{-i \alpha_i} | (2n)_a \rangle \;, \qquad n=0,1,2,.. \;,  \nonumber \\[2mm]
B_k | (2n)_b \rangle &=& e^{i \beta_k} | (2n+1)_b \rangle \;, \qquad B_k | (2n+1)_b \rangle = e^{-i \beta_k} | (2n)_b \rangle \;, \qquad B_k = 1\;\;  \;. \label{ABpair}
\end{eqnarray}
leading to 
\begin{equation}
\langle \eta | {\cal C}_{CHSH} | \eta  \rangle = \frac{ (2 \sqrt{2}) 2 \eta}{1+\eta^2} \;,  \label{max}
\end{equation}
which attains Tsirelson's bound for $\eta \approx 1$. \\\\The representations \eqref{ABpair} have only non-diagonal elements, so that ${\rm Tr}(A_i) = {\rm Tr}(B_k) = 0$. This pattern, already encountered in the finite dimensional case remains valid in the infinite dimesnional case as well. \\\\Let us end this section by evaluating the entanglement entropy in order to understand better the limiting case $\eta \approx 1$. Let us look first at he reduced density matrix $\rho_a$,  obtained from
\begin{equation}
\rho_{ab} = \vert\eta \rangle \langle \eta \vert \;, \label{dens1}
\end{equation}
{\it i.e.}
\begin{equation}
\rho_a = {\rm Tr}_b (\rho_{ab}) = (1-\eta^2) \sum_{n=0}^{\infty} \eta^{2n} \vert n_a \rangle \langle n_a \vert  \;. \label{rho1}
\end{equation}
Therefore
\begin{equation}
\rho_a^2 = (1-\eta^2)^2 \sum_{n=0}^{\infty} \eta^{4n} \vert n_a \rangle \langle n_a \vert  \;, \label{rho2}
\end{equation}
so that
\begin{equation}
{\rm Tr} \rho_a^2 = \frac{1-\eta^2}{1+ \eta^2} \;, \label{trrho}
\end{equation}
from which one sees that $\rho_a$ becomes very impure when $\eta \approx 1$. \\\\Finally, for the  entanglement entropy one gets 
	\begin{equation}
		S = -{\rm Tr} \rho_a \ln \rho_a  = -\ln\left(1-\eta^2\right) - \frac{\eta^2\ln \eta^2}{1-\eta^2} \;. \label{entropy}
	\end{equation}
This entropy is a monotonically increasing function of $\eta$ and diverges for $\eta \rightarrow 1$, confirming that the system is highly entangled in this limit.

\subsection{Relation with the pseudospin operators}

In order to achieve a better understanding of the Bell's operators introduced previously, eqs.\eqref{ABopcoh},\eqref{ABpair}, it is helpful to consider here the so called  pseudospin operators~\cite{psi1,psi2,psi3} defined as
\begin{equation} 
s_x = \sum_{n=0}^\infty s^{(n)}_x \;, \qquad s_y = \sum_{n=0}^\infty s^{(n)}_y \;, \qquad s_z = \sum_{n=0}^\infty s^{(n)}_z \label{spin1} 
\end{equation}
where 
\begin{eqnarray} 
s^{(n)}_x & = & \vert 2n+1 \rangle \langle 2n \vert + \vert 2n \rangle \langle 2n+1 \vert \;, \nonumber \\ 
s^{(n)}_y & = &i\left(  \vert 2n+1 \rangle \langle 2n \vert - \vert 2n \rangle \langle 2n+1 \vert \right) \;, \nonumber \\ 
s^{(n)}_z & = & \vert 2n+1 \rangle \langle 2n+1 \vert - \vert 2n \rangle \langle 2n \vert. \label{spin2}
\end{eqnarray} 
An easy calculation shows that 
\begin{align} 
\left[s^{(n)}_x,s^{(n)}_y \right] &= 2 i s^{(n)}_z \;, \nonumber \\
\left[s^{(n)}_y,s^{(n)}_z \right] &= 2 i s^{(n)}_x \;, \nonumber \\
\left[s^{(n)}_z,s^{(n)}_x \right] &= 2 i s^{(n)}_y.  \label{spin4}
\end{align} 
As a consequence, it follows that the operators~\eqref{spin1} obey the same algebraic relations of the spin $1/2$ Pauli matrices: 
\begin{align} 
\left[ s_x,s_y \right] &= 2 i s_z \;, \nonumber \\
\left[s_y,s_z \right] &= 2 i s_x \;, \nonumber \\
\left[s_z,s_x \right] &= 2 i s_y.  \label{spin5}
\end{align} 
from which the name {\it pseudospin} follows. 

In particular, from expressions \eqref{spin2} one observes that the introduction of the pseudospin operators can be exactly related to the pairing mechanism in Hilbert space, a pair being given by two modes, namely $(\vert2n \rangle, \vert2n+1\rangle)$, with $n=0,1,2,..$. Each pair of modes gives raise to a set of operators, $(s^{(n)}_x,s^{(n)}_y,s^{(n)}_z)$, which obey the spin $1/2$ algebra of the Pauli matrices.  

As already underlined, the observation of the pairing mechanism  goes back to \cite{Gisin}. More recently, its applications to the study of the Bell-CHSH inequality has been discussed in \cite{Sorella:2023hku,Sorella:2023iwz}, where it has been shown that each single pair might be employed for a test of the Bell-CHSH inequality.  

Considering now the first choice, given by equations \eqref{ABopcoh},  it turns out that the operator $A$ can be expressed in terms of the pseudospin operators as 
\begin{equation} 
A = \left( {\vec u} \cdot {\vec s}^{(0)} + {\cal R} \right) \otimes I 
\end{equation}
where $\vec{u}$ denotes the unit vector 
\begin{equation}
{\vec u} =\left( \cos(\alpha), \sin(\alpha),0 \right) \;, \qquad {\vec u} \cdot {\vec u} = 1  \label{vecu}
\end{equation} 
and ${\cal R}$ is the identity operator for $n\ge 2$:
\begin{equation} 
{\cal R} = \sum_{n=2}^\infty \vert n \rangle \langle n \vert    \label{Rop}
\end{equation}
Analogous expressions can be  written down for $B$ as well as for $(A',B')$. For the primed operators, the parameters $\alpha$ and $\beta$ are simply replaced by $\alpha'$ and $\beta'$. 

In the same vein, for the second Bell setup, eq.\eqref{ABpair}, in terms of pseudospin opertors, one has 
\begin{equation} 
A = {\vec u} \cdot {\vec s}   \otimes I   \label{A2st}
\end{equation}
where $\vec{u}$ is the unit vector of expression \eqref{vecu}. 
with similar expressions for the primed operators. \\\\It is immediate to check that in both setups the required properties \eqref{req} for the Bell-type operators are satisfied. \\\\The advantage of employing the pseudospin operators is that they enable to establish a simple and clear relationship between the various inequivalent representations of the Bell operators and the algebra of the Pauli matrices \cite{VS}. \\\\Remark that, from equation  \eqref{A2st},  it follows that 
\begin{equation} 
{\rm Tr}A =0 \;, \label{once} 
\end{equation} 
a property already mentioned and to which we shall devote the next considerations. 

\section{The traceless representation} 

In all examples treated so far, the maximum violation of the Bell-CHSH occurs for the representation for which the Bell operators are traceless, namely {\rm Tr}A =0, see equations  \eqref{opAB24}, \eqref{sixmatr}, \eqref{A2st}. Such a representation is obtained by the direct sum of the 2x2 matrix ${\cal M}_{i} $ of equations 
\eqref{MB}, \eqref{nsigma}. In fact, equations  \eqref{opAB24}, \eqref{sixmatr}, \eqref{A2st} can be written, respectively, as 
\begin{eqnarray} 
& {\cal M}_{i}   \oplus {\cal M}_{i}  \nonumber \\
& {\cal M}_{i}  \oplus {\cal M}_{i} \oplus {\cal M}_{i} \nonumber \\
& {\cal M}_{i}  \oplus {\cal M}_{i} \oplus {\cal M}_{i}  \oplus {\cal M}_{i} \oplus ...........   \label{oplus} 
\end{eqnarray} 
In practice, in the case of the maximum violation off the Bel-CHSH inequality we have essentially many replica of the elementary spin $1/2$ result, in agreement with \cite{VS}.

\section{Conclusion}\label{conclus}
In the present work we have pointed out that, when the dimension of the Hilbert space is greater than two, the Bell operators, eq.\eqref{req}, entering the Bell-CHSH inequality exhibit unitarily inequivalent representations, leading to different values of the violation. \\\\This feature  can be traced back to the two by two traceless matrix ${\cal M}_{i}$ 
\begin{equation}
{\cal M}_{i} = {\cal M}_{i}^{\dagger } = 
\begin{pmatrix}
0 & e^{i \alpha_i}  \\
e^{-i \alpha_i}  & 0  \\
\end{pmatrix} 
\;, \label{MBconc}
\end{equation} 
which turns out to be precisley Bell's operator for spin $1/2$. As discussed in the previous sections, the role of 
${\cal M}_{i}$, which is the building block of the varius representations, is that of organizing the modes entering a given entangled state $| \psi \rangle$ into pairs, eqs.\eqref{pair},\eqref{p2pair}. \\\\When the matrix ${\cal M}_{i}$ appears only once, the violation of the Bell-CHSH inequality is the lowest possible. However, when all modes are grouped into pairs, something which can happen only when the dimension of the Hilbert space is even, the violation is the biggest one, attaining Tsirelson bound: $2 \sqrt{2}$, \\\\For odd dimensional Hilbert spaces, Tsirelson's bound is never achieved. \\\\Everything generalizes to infinite dimensional Hilbert spaces.

\section*{Acknowledgements}
The authors would like to thank the Brazilian agencies CNPq and FAPERJ for financial support.  S.P.~Sorella is a level $1$ CNPq researcher under the contract 301030/2019-7.

\end{document}